\documentstyle[12pt]{article}
\def\a{\alpha}
\def\b{\beta}
\def\m{\mu}
\def\tm{\tilde{m}}
\def\th{\theta}
\def\tr{{\rm Tr}}
\def\pf{{\rm Pf}}

\def\de{\delta}
\def\P{\Phi}

\def\l{\lambda}
\def\L{\Lambda}

\def\e{\epsilon}

\def\t{\tau}
\def\G{\Gamma}
\def\Z{{\bf Z}}
\def\implies{\Rightarrow}

\begin{document}

\newcommand{\inv}[1]{{#1}^{-1}} %inverse

\renewcommand{\theequation}{\thesection.\arabic{equation}}
\newcommand{\beq}{\begin{equation}}
\newcommand{\eeq}[1]{\label{#1}\end{equation}}
\newcommand{\ber}{\begin{eqnarray}}
\newcommand{\eer}[1]{\label{#1}\end{eqnarray}}
\begin{titlepage}
\begin{center}
% 4 December, 1995
                                \hfill    CERN-TH/95-329 RI-12-95\\
                                \hfill    hep-th/9512140\\

\vskip .5in

{\large \bf Summary of Results in $N=1$ Supersymmetric $SU(2)$ Gauge
Theories}~\footnotemark
\footnotetext{Talk presented by Amit Giveon at the {\em 29th International
Symposium on the Theory of Elementary Particles} in Buckow, Germany,
August 29 - September 2, 1995, and at the workshop on {\em STU-Dualities and
Non-Perturbative Phenomena in Superstrings and Supergravity}, CERN, Geneva,
November 27 - December 1, 1995; to appear in the proceedings.}
\vskip .4in

{\large  S. Elitzur},$^a$ \footnotemark\
\footnotetext{e-mail address: elitzur@vms.huji.ac.il}
{\large  A. Forge},$^a$ \footnotemark\
\footnotetext{e-mail address: forge@vms.huji.ac.il}
{\large  A. Giveon},$^b$ \footnotemark\
\footnotetext{On leave of absence from Racah Institute of Physics, The Hebrew
University, Jerusalem 91904, Israel; e-mail address: giveon@vxcern.cern.ch}
{\large  E. Rabinovici}$^a$ \footnotemark\
\footnotetext{e-mail address: eliezer@vms.huji.ac.il}
\vskip .3in

$a$
{\em Racah Institute of Physics, The Hebrew University\\
  Jerusalem, 91904, Israel} \\
\vskip .1in
$b$
{\em Theory Division, CERN, CH 1211, Geneva 23, Switzerland}\\

\vskip .1in
\end{center}
\vskip .2in
\begin{center} {\bf ABSTRACT } \end{center}
\begin{quotation}
\noindent

We summarize some results in $4d$, $N=1$
supersymmetric $SU(2)$ gauge theories: the exact effective superpotentials,
the vacuum structure, and the exact effective Abelian couplings
for arbitrary bare masses and Yukawa couplings.

\end{quotation}

\vfill

\begin{flushleft}
CERN-TH/95-329\\
December, 1995
\end{flushleft}

\end{titlepage}
\eject
\def\baselinestretch{1.2}
\baselineskip 16 pt

\noindent
\section{The Main Result}
\setcounter{equation}{0}

The new results in this summary are based on some of the results in refs.
\cite{efgr,efgr2}.
We consider $N=1$ supersymmetric $SU(2)$ gauge theories in four dimensions,
with any possible content of matter superfields, such that the theory is
either one-loop asymptotic free or conformal. This allows the introduction of
$2N_f$ matter supermultiplets in the fundamental representation,
$Q_i^a$, $i=1,...,2N_f$,
$N_A$ supermultiplets in the adjoint representation,
$\P_{\a}^{ab}$, $\a=1,...,N_A$,
and $N_{3/2}$ supermultiplets in the spin 3/2 representation, $\Psi$.
Here $a,b$ are fundamental representation indices, and $\P^{ab}=\P^{ba}$
(we present $\Psi$ in a schematic form as we shall not use it much). The
numbers $N_f$, $N_A$ and $N_{3/2}$ are limited by the condition:
\beq
b_1=6-N_f-2N_A-5N_{3/2}\geq 0,
\eeq{b1}
where $-b_1$ is the one-loop coefficient of the gauge coupling beta-function.

The main result of this talk is the following: the effective
superpotential of an (asymptotically free or conformal)
$N=1$ supersymmetric $SU(2)$ gauge theory,
with $2N_f$ doublets and $N_A$ triplets (and $N_{3/2}$ quartets) is
\ber
W_{N_f,N_A}(M,X,Z,N_{3/2}) &=& -\de_{N_{3/2},0}
(4-b_1)\Big\{\L^{-b_1} \pf_{2N_f} X\Big[
{\rm det}_{N_A}(\Gamma_{\a\b})\Big]^2
\Big\}^{1/(4-b_1)}\nonumber\\
&+&\tr_{N_A} \tm M +{1\over 2}\tr_{2N_f} mX
+{1\over\sqrt{2}}\tr_{2N_f} \l^{\a} Z_{\a}+\de_{N_{3/2},1} gU ,
\eer{W}
where
\beq
\Gamma_{\a\b}(M,X,Z)=M_{\a\b}+\tr_{2N_f}(Z_{\a}X^{-1}Z_{\b}X^{-1}).
\eeq{G}
The first term in (\ref{W}) is the exact (dynamically generated)
nonperturbative superpotential\footnote{Integrating in the ``glueball'' field
$S=-W_{\a}^2$, whose source is $\log \L^{b_1}$, gives the nonperturbative
superpotential:
$$ W(S,M,X,Z)=S\Big[\log \Big({\L^{b_1}S^{4-b_1}\over
\pf X({\rm det}\Gamma )^2}\Big)-(4-b_1)\Big] .$$}, and the other terms are the
tree-level superpotential.
$\L$ is the dynamically generated scale, while $\tm_{\a\b}$, $m_{ij}$
and $\l^{\a}_{ij}$ are the bare masses and Yukawa couplings, respectively
($\tm_{\a\b}=\tm_{\b\a}$, $m_{ij}=-m_{ji}$,
$\l^{\a}_{ij}=\l^{\a}_{ji}$). The gauge singlets, $X$, $M$, $Z$, $U$, are
given in terms of the $N=1$ superfield doublets, $Q^a$, the triplets,
$\P^{ab}$, and the quartets, $\Psi$,  as follows:
\ber
X_{ij}&=&Q_{ia} Q_j^a, \qquad  a=1,2, \qquad i,j=1,...,2N_f,
\nonumber\\
M_{\a\b}&=&\P_{\a b}^a\P_{\b a}^b, \qquad \a ,\b=1,...,N_A, \qquad a,b=1,2,
\nonumber\\
Z_{ij}^{\a}&=&Q_{ia}\P_{\a b}^a Q_j^b , \qquad \qquad U=\Psi^4.
\eer{XMZ}
Here, the $a,b$ indices are raised and lowered with an $\e_{ab}$ tensor.
The gauge-invariant superfields, $X_{ij}$, may be considered as a mixture of
$SU(2)$ ``mesons'' and ``baryons'', while the gauge-invariant superfields,
$Z_{ij}^{\a}$, may be considered as a mixture of $SU(2)$ ``meson-like'' and
``baryon-like'' operators.

Equation (\ref{W}) is a universal representation of the superpotential for all
infra-red non-trivial theories; all the physics we shall discuss (and beyond)
is in (\ref{W}). In particular, all the symmetries and quantum numbers of the
various parameters are already embodied in $W_{N_f,N_A}$.
The nonperturbative superpotential is derived in refs. \cite{efgr,efgr2} by an
``integrating in'' procedure, following refs. \cite{ILS,I}. The details can be
found in ref. \cite{efgr2}, and will not be presented here. Instead, we list
the main results concerning each of the theories, $N_f,N_A,N_{3/2}$,
case by case.

\noindent
\section{$b_1=6$: $N_f=N_A=N_{3/2}=0$}
\setcounter{equation}{0}

This is a pure $N=1$ supersymmetric $SU(2)$ gauge theory. The nonperturbative
effective superpotential is\footnote{This can be read from eq. (\ref{W}) by
setting $\tr_{2N_f}(\cdot)=0$, $\det_{2N_f}(\cdot)=1$
(for example, $\pf X=1$, $\G=M$) when $N_f=0$, and $\tr_{N_A}(\cdot)=0$,
$\det_{N_A}(\cdot)=1$ (for example, $\det\Gamma =1$) when $N_A=0$; this will
also be used later.}
\beq
W_{0,0}=\pm 2 \L^3 .
\eeq{61}
This theory was considered before \cite{lec}. The superpotential in
eq. (\ref{61}) can be derived by integrating out the matter of any of the
other theories; it is non-zero due to gluino condensation. The ``$\pm$'' in
(\ref{61}) comes from the square-root, appearing on the braces in (\ref{W}),
when $b_1=6$; it corresponds, physically, to the two quantum vacua of a
pure $N=1$ supersymmetric $SU(2)$ gauge theory.

\noindent
\section{$b_1=5$: $N_f=1$, $N_A=N_{3/2}=0$}
\setcounter{equation}{0}
There is one case with $b_1=5$, namely, $SU(2)$ with one flavor. This theory
was considered before \cite{lec}; it is a particular case of $SU(N_c)$
with $N_f=N_c-1$. The superpotential is
\beq
W_{1,0}={\L^5\over X}+mX ,
\eeq{51}
where $X$ and $m$ are defined by: $X_{ij}\equiv X\e_{ij}$,
$m_{ij}\equiv -m\e_{ij}$. The nonperturbative part of $W_{1,0}$ is proportional
to the one instanton action. The vacuum degeneracy of the classical low-energy
effective theory is lifted quantum mechanically; from eq. (\ref{51})
we see that, in the massless case, there is no vacuum at all.

\noindent
\section{$b_1=4$}
\setcounter{equation}{0}
There are two cases with $b_1=4$: either $N_f=2$, or $N_A=1$.
In both cases, the nonperturbative superpotential vanishes
and, in addition, there is a
constraint\footnote{This is reflected in eq. (\ref{W}) by the vanishing of the
coefficient $(4-b_1)$ in front of the braces, leading to $W=0$, and the
singular power $1/(4-b_1)$ on the braces,  when $b_1=4$, which signals the
existence of a constraint.}.

\noindent
\subsection{$N_f=2$, $N_A=N_{3/2}=0$}
The nonperturbative superpotential vanishes
\beq
W_{2,0}^{non-per.}=0 ,
\eeq{41}
and by the integrating in procedure we also get the quantum constraint:
\beq
\pf X=\L^4 .
\eeq{42}
This theory was considered before \cite{lec}; it is a particular
case of $SU(N_c)$ with $N_f=N_c$.
At the classical limit, $\L\to 0$, the quantum constraint collapses into the
classical constraint, $\pf X=0$.

\noindent
\subsection{$N_f=0$, $N_A=1$, $N_{3/2}=0$}
The massless $N_A=1$ case is a pure $SU(2)$, $N=2$ supersymmetric Yang-Mills
theory. This model was considered in detail in ref. \cite{SW1}.
The nonperturbative superpotential vanishes
\beq
W_{0,1}^{non-per.}=0 ,
\eeq{43}
and by the integrating in procedure we also get the quantum constraint:
\beq
M=\pm \L^2 .
\eeq{44}
This result can be understood because the starting point of the integrating
in procedure is a pure $N=1$ supersymmetric Yang-Mills theory. Therefore,
it leads us to the points at the verge of confinement in the moduli space.
These are the two singular points in the $M$ moduli space of the theory;
they are due to massless monopoles or dyons. Such excitations are not
constructed out of the elementary degrees of freedom and, therefore, there
is no trace for them in $W$. (This situation is different if $N_f\neq 0$,
$N_A=1$; in this case, monopoles are different manifestations of the
elementary degrees of freedom.)

\noindent
\section{$b_1=3$}
\setcounter{equation}{0}
There are two cases with $b_1=3$: either $N_f=3$, or $N_A=N_f=1$. In both
cases, for vanishing bare parameters in (\ref{W}), the semi-classical limit,
$\L\to 0$, imposes the classical constraints, given by the equations of
motion: $\partial W=0$; however, quantum corrections remove the constraints.

\noindent
\subsection{$N_f=3$, $N_A=N_{3/2}=0$}
The superpotential is
\beq
W_{3,0}=-{\pf X\over \L^3}+{1\over 2}\tr mX .
\eeq{31}
This theory was considered before \cite{lec}; it is a particular
case of $SU(N_c)$ with $N_f=N_c+1$.
In the massless case, the equations $\partial_X W=0$ give the classical
constraints; in particular, the superpotential is proportional to a classical
constraint: $\pf X=0$. The negative power of $\L$, in eq. (\ref{31}) with
$m=0$, indicates
that small values of $\L$ imply a semi-classical limit for which the classical
constraints are imposed.

\noindent
\subsection{$N_f=1$, $N_A=1$, $N_{3/2}=0$}
In this case, the superpotential in (\ref{W}) reads
\beq
W_{1,1}=-{\pf X\over \L^3}\G^2+\tm M +{1\over 2}\tr mX + {1\over \sqrt{2}}\tr
\l Z .
\eeq{32}
Here $m$, $X$ are antisymmetric $2\times 2$ matrices, $\l$, $Z$ are
symmetric $2\times 2$ matrices and
\beq
\G=M+\tr (ZX^{-1})^2 .
\eeq{33}
This superpotential was found before in ref. \cite{IS1}. To find the
quantum vacua, we solve the equations: $\partial_M W =\partial_X W
=\partial_Z W =0$. Let us discuss some properties of this theory:

\begin{itemize}

\item
The equations $\partial W=0$ can be re-organized into the singularity
conditions of an elliptic curve:
\beq
y^2=x^3+ax^2+bx+c
\eeq{34}
(and some other equations), where the coefficients $a,b,c$ are functions of
only the field $M$, the scale $\L$, the bare quark masses, $m$, and Yukawa
couplings, $\l$. Explicitly,
\beq
a=-M, \qquad b={\L^3\over 4}\pf m, \qquad c=-{\a\over 16} ,
\eeq{35}
where
\beq
\a={\L^6\over 4}\det\l .
\eeq{36}

\item
The parameter $x$, in the elliptic curve (\ref{34}), is given in terms of the
composite field:
\beq
x\equiv {1\over 2}\G .
\eeq{37}

\item
$W_{1,1}$ has $2+N_f=3$ vacua, namely, the  three singularities of the
elliptic curve in (\ref{34}), (\ref{35}). These are the three solutions,
$M(x)$, of the equations: $y^2=\partial y^2/\partial x=0$; the solutions
for $X$, $Z$ are given by the other equations of motion.

\item
The {\em 3 quantum vacua} are the vacua of the theory in the
{\em Higgs-confinement phase} \cite{BR}.

\item
Phase transition points to the {\em Coulomb branch} are at
$X=0 \Leftrightarrow \tm=0$.
Therefore, we conclude that the elliptic curve defines the {\em effective
Abelian coupling}, $\tau(M,\L,m,\l)$, in the Coulomb branch.

\item
On the subspace of bare parameters, where the theory has an enhanced $N=2$
supersymmetry, the result in eq. (\ref{35}) coincides with the result in
\cite{SW2} for $N_f=1$.

\item
In the massless case, there is a $Z_{4-N_f}=Z_3$ global symmetry acting on the
moduli space.

\item
When the masses and Yukawa couplings approach zero, all the 3 singularities
collapse to the origin. Such a point might be interpreted as a
new scale-invariant theory \cite{lec}. As before, the
negative power of $\L$, in eq. (\ref{32}) with $\tm=m=\l=0$,
indicates that small values of
$\L$ imply a semi-classical limit for which the classical constraint, $\G=0$,
is imposed. Indeed, for vanishing bare parameters, the equations of motion
are solved by any $M,X,Z$ obeying $\G=0$.

\end{itemize}

\noindent
\section{$b_1=2$}
\setcounter{equation}{0}
There are three cases with $b_1=2$: $N_f=4$, or $N_A=1$, $N_f=2$, or $N_A=2$.
In all three cases, for vanishing bare parameters in (\ref{W}), there
are extra massless degrees of freedom not included in the procedure; those are
expected due to a non-Abelian conformal theory.

\noindent
\subsection{$N_f=4$, $N_A=N_{3/2}=0$}
The superpotential is
\beq
W_{4,0}=-2{(\pf X)^{{1\over 2}}\over \L}+{1\over 2}\tr mX .
\eeq{21}
This theory was considered before in \cite{lec}; it is a particular
case of $SU(N_c)$ with $N_f>N_c+1$.
In the massless case, the superpotential is proportional to the square-root of
a classical constraint: $\pf X=0$. The branch cut at $\pf X=0$ signals the
appearance of extra massless degrees of freedom at these points; those are
expected in ref. \cite{S2}.
Therefore, we make use of the superpotential only in the presence of masses,
$m$, which fix the vacua away from such points.

\noindent
\subsection{$N_f=2$, $N_A=1$, $N_{3/2}=0$}
In this case, the superpotential in (\ref{W}) reads
\beq
W_{2,1}=-2{(\pf X)^{{1\over 2}}\over \L}\G
+\tm M +{1\over 2}\tr mX + {1\over \sqrt{2}}\tr \l Z .
\eeq{22}
Here $m$, $X$ are antisymmetric $4\times 4$ matrices, $\l$, $Z$ are
symmetric $4\times 4$ matrices and $\G$ is given in eq. (\ref{33}).
As in section 5.2, to find the quantum vacua, we solve the equations:
$\partial W =0$. Let us discuss some properties of this theory:

\begin{itemize}

\item
The equations $\partial W=0$ can be re-organized into the singularity
conditions of an elliptic curve (\ref{34}) (and some other equations), where
the coefficients $a,b,c$ are functions of
only the field $M$, the scale $\L$ and the bare quark masses, $m$, and Yukawa
couplings, $\l$. Explicitly \cite{efgr,efgr2},
\beq
a= -M, \qquad b= -{\a\over 4}+{\L^2\over 4}\pf m , \qquad
c= {\a\over 8}\Big(2M+\tr(\m^2)\Big),
\eeq{23}
where
\beq
\a={\L^4\over 16}\det\l , \qquad \m=\l^{-1}m .
\eeq{24}

\item
As in section 5.2, the parameter $x$, in the elliptic curve (\ref{34}),
is given in terms of the composite field:
\beq
x\equiv {1\over 2}\G .
\eeq{25}
Therefore, we have identified a physical meaning of the parameter $x$.

\item
$W_{2,1}$ has $2+N_f=4$ vacua, namely, the four singularities of the
elliptic curve in (\ref{34}), (\ref{23}). These are the four solutions,
$M(x)$, of the equations: $y^2=\partial y^2/\partial x=0$; the solutions
for $X$, $Z$ are given by the other equations of motion.

\item
The {\em 4 quantum vacua} are the vacua of the theory in the
{\em Higgs-confinement phase}.

\item
Phase transition points to the {\em Coulomb branch} are at
$X=0 \implies \tm=0$.
Therefore, we conclude that the elliptic curve defines the {\em effective
Abelian coupling}, $\tau(M,\L,m,\l)$, in the Coulomb branch.

\item
On the subspace of bare parameters, where the theory has an enhanced $N=2$
supersymmetry, the result in eq. (\ref{23}) coincides with the result in
\cite{SW2} for $N_f=2$.

\item
In the massless case, there is a $Z_{4-N_f}=Z_2$ global symmetry acting on the
moduli space.

\item
For special values of the bare masses and Yukawa couplings, some of the 4 vacua
degenerate. In some cases, it may lead to points where mutually non-local
degrees of freedom are massless, similar to the situation in pure $N=2$
supersymmetric gauge theories, considered in \cite{AD}. For example, when the
masses and Yukawa couplings approach zero, all the 4 singularities
collapse to the origin. Such points might be interpreted as in a
{\em non-Abelian Coulomb phase} \cite{lec}.

\item
The singularity at $X=0$ (in $\G$) and the branch cut at $\pf X=0$
(due to the $1/2$ power in eq. (\ref{22}))
signal the appearance of extra massless degrees of freedom at these points;
those are expected similar to refs. \cite{S2,K}.
Therefore, we make use of the superpotential only in the presence of bare
parameters, which fix the vacua away from such points.

\end{itemize}

\noindent
\subsection{$N_f=0$, $N_A=2$, $N_{3/2}=0$}
In this case, the superpotential in eq. (\ref{W}) reads
\beq
W_{0,2}=\pm 2{\det M\over \L}+\tr \tm M .
\eeq{26}
Here $\tm$, $M$ are $2\times 2$ symmetric matrices, and the ``$\pm$'' comes
from the square-root, appearing on the braces in (\ref{W}), when $b_1=2$.
The superpotential in eq. (\ref{26}) is the one presented in \cite{IS1,IS2}
on the confinement and the oblique confinement branches\footnote{
The fractional power $1/(4-b_1)$ on the braces
in (\ref{W}), for any theory with $b_1\leq 2$, may indicate a similar
phenomenon, namely, the existence of confinement and oblique
confinement branches of the theory, corresponding to the $4-b_1$
phases due to the fractional power. It is plausible that, for $SU(2)$, such
branches are related by a discrete symmetry.}
(they are related by a discrete symmetry \cite{lec}). This theory has
{\em two quantum vacua}; these become the phase transition points to the
Coulomb branch when $\det \tm =0$. The moduli space may also contain a
non-Abelian Coulomb phase when the two singularities degenerate at $M=0$
\cite{IS1}; this happens when $\tm=0$. At this point, the theory has extra
massless degrees of freedom and, therefore, $W_{0,2}$ fails to describe the
physics at $\tm=0$. Moreover, at $\tm=0$, the theory has other descriptions
via an electric-magnetic triality \cite{lec}.

\noindent
\section{$b_1=1$}
\setcounter{equation}{0}
There are four cases with $b_1=1$: $N_f=5$, or $N_A=1$, $N_f=3$, or $N_A=2$,
$N_f=1$, or $N_{3/2}=1$.

\noindent
\subsection{$N_f=5$, $N_A=N_{3/2}=0$}
The superpotential is
\beq
W_{5,0}=-3{(\pf X)^{{1\over 3}}\over \L^{{1\over 3}}}+{1\over 2}\tr mX .
\eeq{11}
This theory was considered before in \cite{lec}; it is a particular
case of $SU(N_c)$ with $N_f>N_c+1$. The discussion in section 6.1 is
relevant in this case too.

\noindent
\subsection{$N_f=3$, $N_A=1$, $N_{3/2}=0$}
In this case, the superpotential in (\ref{W}) reads
\beq
W_{3,1}=-3{(\pf X)^{{1\over 3}}\over \L^{{1\over 3}}}\G^{{2\over 3}}
+\tm M +{1\over 2}\tr mX + {1\over \sqrt{2}}\tr \l Z .
\eeq{12}
Here $m$, $X$ are antisymmetric $6\times 6$ matrices, $\l$, $Z$ are
symmetric $6\times 6$ matrices and $\G$ is given in eq. (\ref{33}).
Let us discuss some properties of this theory:

\begin{itemize}

\item
The equations $\partial W=0$ can be re-organized into the singularity
conditions of an elliptic curve (\ref{34}) (and some other equations), where
the coefficients $a,b,c$ are \cite{efgr,efgr2}
\ber
a&=& -M-\a , \qquad b\,\, =\,\, 2\a  M +
{\a\over 2}\tr(\m^2) + {\L\over 4} \pf m , \nonumber\\
c&=&{\a\over 8}\Big( -8M^2-4M \tr(\m^2) - [\tr(\m^2)]^2 + 2\tr(\m^4) \Big) ,
\eer{13}
where
\beq
\a={\L^2\over 64}\det\l , \qquad \m=\l^{-1}m .
\eeq{14}
In eq. (\ref{13}) we have shifted the quantum field $M$ to
\beq
M\to M-\a/4 .
\eeq{15}

\item
The parameter $x$, in the elliptic curve (\ref{34}), is given in terms of the
composite field:
\beq
x\equiv {1\over 2}\G + {\a\over 2}.
\eeq{16}
Therefore, as before, we have identified a physical meaning of the parameter
$x$.

\item
$W_{3,1}$ has $2+N_f=5$ {\em quantum vacua}, corresponding to the 5
singularities of the elliptic curve (\ref{34}), (\ref{13});
these are the vacua of the theory in the {\em Higgs-confinement phase}.

\item
{}From the phase transition points to the {\em Coulomb branch}, we conclude
that
the elliptic curve defines the {\em effective Abelian coupling},
$\tau(M,\L,m,\l)$, for arbitrary bare masses and Yukawa couplings.
As before, on the subspace of bare parameters, where the theory has $N=2$
supersymmetry, the result in eq. (\ref{13}) coincides with the result in
\cite{SW2} for $N_f=3$.

\item
For special values of the bare masses and Yukawa couplings, some of the 5 vacua
degenerate. In some cases, it may lead to points where mutually non-local
degrees of freedom are massless, and might be interpreted as in a
{\em non-Abelian Coulomb phase} or another
new {\em superconformal theory} in four
dimensions (see the discussion in section 6.2).

\item
The singularity  and branch cuts in $W_{3,1}$
signal the appearance of extra massless degrees of freedom at these points.
The discussion in the end of section 6.2 is relevant here too.

\end{itemize}

\noindent
\subsection{$N_f=1$, $N_A=2$, $N_{3/2}=0$}
In this case, the superpotential in (\ref{W}) reads \cite{efgr2}
\beq
W_{1,2}=-3{(\pf X)^{1/3}\over \L^{1/3}}(\det\G)^{2/3}+\tr \tm M
+{1\over 2}\tr mX + {1\over \sqrt{2}} \tr \l^{\a} Z_{\a}.
\eeq{17}
Here $m$ and $X$ are  antisymmetric $2\times 2$ matrices,
$\l^{\a}$ and $Z_{\a}$ are symmetric $2\times 2$ matrices, $\a=1,2$,
$\tm$, $M$ are $2\times 2$ symmetric matrices and $\G_{\a\b}$ is given in eq.
(\ref{G}).
This theory has {\em 3 quantum vacua} in the Higgs-confinement branch. At the
phase transition points to the Coulomb branch, namely, when $\det\tm =0
\Leftrightarrow \det M=0$, the equations of motion can be re-organized into the
singularity conditions of an elliptic curve (\ref{34}). Explicitly, when
$\tm_{22}=\tm_{12}=0$, the coefficients $a,b,c$ in (\ref{34}) are \cite{efgr2}
\beq
a=-M_{22},\qquad b={\L\tm_{11}^2\over 16}\pf m,\qquad
c=-\Big({\L\tm_{11}^2\over 32}\Big)^2 \det\l_2 .
\eeq{18}
However, unlike the $N_A=1$ cases, the equations $\partial W=0$ {\em cannot} be
re-organized into the singularity condition of an elliptic curve, in general.
This result makes sense, physically, since an elliptic curve is expected to
``show up'' only at the phase transition points to the Coulomb branch.
For special values of the bare parameters, there are points in the moduli
space where (some of) the singularities degenerate; such points might be
interpreted as in a non-Abelian Coulomb phase, or new superconformal theories.
For more details, see ref. \cite{efgr2}.

\noindent
\subsection{$N_f=N_A=0$, $N_{3/2}=1$}
This chiral theory was shown to have $W_{0,0}^{non-per.}(N_{3/2}=1)=0$
\cite{ISS}; perturbing it by a tree-level superpotential, $W_{tree}=gU$, where
$U$ is given in (\ref{XMZ}), may
lead to dynamical supersymmetry breaking \cite{ISS}.

\noindent
\section{$b_1=0$}
\setcounter{equation}{0}
There are five cases with $b_1=0$: $N_f=6$, or $N_A=1$, $N_f=4$, or
$N_A=N_f=2$, or $N_A=3$, or $N_{3/2}=N_f=1$. These theories have vanishing
one-loop beta-functions in either conformal or infra-red free beta-functions
and, therefore, will possess extra structure.

\noindent
\subsection{$N_f=6$, $N_A=N_{3/2}=0$}
This theory is a particular case of $SU(N_c)$ with $N_f=3N_c$; the electric
theory is free in the infra-red \cite{lec}.\footnote{
A related fact is that
(unlike the $N_A=1$, $N_f=4$ case, considered next) in the (would be)
superpotential, $W_{6,0}=-4\L^{-b_1/4}(\pf X)^{1/4}+{1\over 2}\tr mX$,
it is impossible to construct the matching ``$\L^{b_1}$''$\equiv f(\t_0)$,
where $\t_0$ is the non-Abelian gauge coupling constant,
in a way that respects the global symmetries.}

\noindent
\subsection{$N_f=4$, $N_A=1$, $N_{3/2}=0$}
In this case, the superpotential in (\ref{W}) reads
\beq
W_{4,1}=-4{(\pf X)^{{1\over 4}}\over \L^{{b_1\over 4}}}\G^{{1\over 2}}
+\tm M +{1\over 2}\tr mX + {1\over \sqrt{2}}\tr \l Z .
\eeq{01}
Here $m$, $X$ are antisymmetric $8\times 8$ matrices, $\l$, $Z$ are
symmetric $8\times 8$ matrices, $\G$ is given in eq. (\ref{33}) and
\beq
\L^{b_1}\equiv 16\a(\t_0)^{1/2}(\det \l)^{-1/2} ,
\eeq{02}
where $\a(\t_0)$ will be presented in eq. (\ref{04}).
Let us discuss some properties of this theory:

\begin{itemize}

\item
The equations $\partial W=0$ can be re-organized into the singularity
conditions of an elliptic curve (\ref{34}) (and some other equations), where
the coefficients $a,b,c$ are \cite{efgr,efgr2}
\ber
a&=&{1\over \b^2}\Big\{
2{\a+1\over \a-1}M+{8\over \b^2}{\a\over (\a-1)^2}\tr(\m^2)\Big\}, \nonumber\\
b&=&{1\over \b^4}\Big\{
-16{\a\over (\a-1)^2} M^2 + {32\over \b^2}{\a(\a+1)\over (\a-1)^3}M\tr(\m^2)
\nonumber\\
&-&{8\over \b^4}{\a\over (\a-1)^2}\Big[(\tr (\m^2))^2-2\tr(\m^4)\Big]+
{4\over \b^4}{(\a+1)\L^{b_1}\over (\a-1)^2} \pf m\Big\} , \nonumber\\
c&=& {1\over \b^6}\Big\{
-32{\a(\a+1)\over (\a-1)^3}M^3+{32\over \b^2}{\a(\a+1)^2\over
(\a-1)^4}M^2\tr(\m^2)\nonumber\\
&+&M\Big[-{16\over \b^4}{\a(\a+1)\over (\a-1)^3}
\Big((\tr(\m^2))^2-2\tr(\m^4)\Big) + {32\over \b^4}{\a\L^{b_1}\over (\a-1)^3}
\pf m \Big]\nonumber\\
&-&{32\over \b^6}{\a\over (\a-1)^2}\Big[\tr(\m^2)\tr(\m^4)-{1\over
6}(\tr(\m^2))^3-{4\over 3}\tr(\m^6)\Big]\Big\}.
\eer{03}
Here $\m=\l^{-1}m$ and $\a$, $\b$ are functions of $\tau_0$, the non-Abelian
gauge coupling constant; comparison with ref. \cite{SW2} gives
\beq
\a(\tau_0)\equiv {``\L^{2b_1}"\over 256} \det\l
=\left({\th_2^2-\th_3^2\over \th_2^2+\th_3^2}\right)^2, \qquad
\b(\tau_0)={\sqrt{2}\over \th_2\th_3},
\eeq{04}
where
\beq
\th_2(\tau_0)=\sum_{n\in Z}(-1)^n e^{\pi i \tau_0 n^2}, \qquad
\th_3(\tau_0)=\sum_{n\in Z}e^{\pi i \tau_0 n^2}, \qquad
\tau_0={\th_0\over \pi}+{8\pi i\over g_0^2}.
\eeq{05}
In eq. (\ref{03}) we have shifted the quantum field $M$ to
\beq
M\to \b^2 M-\a\tr\m^2/(\a-1) .
\eeq{06}

\item
The parameter $x$, in the elliptic curve (\ref{34}), is given in terms of
the composite field:
\beq
x\equiv {1\over \b^4}\Big[\G - {4\a\over (\a-1)^2}\tr\m^2\Big].
\eeq{07}

\item
$W_{4,1}$ has $2+N_f=6$ {\em quantum vacua}, corresponding to the 6
singularities of the elliptic curve (\ref{34}), (\ref{03}); these are the
vacua of the theory in the {\em Higgs-confinement phase}.

\item
As before, from the phase transition points to the {\em Coulomb branch}, we
conclude that the elliptic curve defines the {\em effective Abelian coupling},
$\tau(M,\L,m,\l)$, for arbitrary bare masses and Yukawa couplings.
On the subspace of bare parameters, where the theory has $N=2$
supersymmetry, the result in eq. (\ref{03}) coincides with the result in
\cite{SW2} for $N_f=4$.

\item
The discussion in the end of sections 6.2 and 7.2 is relevant here too.

\item
We can get the results for $N_A=1$, $N_f<4$, by integrating out flavors.

\item
In all the $N_A=1$, $N_f\neq 0$ cases we {\em derived} the result that $\t$ is
a section of an $SL(2,\Z)$ bundle over the moduli space and over the parameters
space of bare masses and Yukawa couplings (since $\t$ is a modular parameter of
a torus).

\end{itemize}

\noindent
\subsection{$N_f=2$, $N_A=2$, $N_{3/2}=0$}
It was argued that this theory is infra-red free \cite{efgr2}.\footnote{A
related fact is that (unlike the $N_A=1$, $N_f=4$ case) it is impossible to
construct a matching, ``$\L^{b_1}$''$\equiv\a(\t_0)f(\l^{\a})$, in a way
that respects the global symmetries.}

\noindent
\subsection{$N_f=0$, $N_A=3$, $N_{3/2}=0$}
In this case, the superpotential in eq. (\ref{W}) reads
\beq
W_{0,3}=-4{(\det M)^{{1\over 2}}\over \L^{{b_1\over 4}}}+\tr \tm M .
\eeq{08}
Here $\tm$, $M$ are $3\times 3$ symmetric matrices. The superpotential
(\ref{08}) equals to the  tree-level superpotential, $W_{tree}=\l\det\P$,
where, schematically,  $\det\P\sim \e\P\P\P\sim (\det M)^{1/2}$ is the
(antisymmetric) coupling of the three triplets, $\P_{\a}$.
This result coincides
with the one derived in \cite{IS2}. Therefore, we identify the matching
``$\L^{-b_1/4}$''$\equiv \l f(\t_0)$, which respects the global symmetries.
In the massless case, this theory flows to an $N=4$ supersymmetric Yang-Mills
fixed point.

\noindent
\subsection{$N_f=1$, $N_A=0$, $N_{3/2}=1$}
It was argued that this theory is infra-red free \cite{efgr}.

\noindent
\section{More Results}
\setcounter{equation}{0}

We have summarized some old and new results in $N=1$ supersymmetric $SU(2)$
gauge theories. More new results, along the lines of this investigation,
were derived in \cite{efgr2,efgr3,efgr4}. In ref. \cite{efgr2}, we have
derived some results in $N=1$ supersymmetric $SU(N_c)$ gauge theories,
$N_c>2$, with $N_A$ matter supermultiplets in the adjoint representation,
$N_f$ supermultiplets in the fundamental representation and $N_f$
supermultiplets in the anti-fundamental representation. More properties
of $SU(N_c)$ supersymmetric gauge theories were studied and will be reported
in \cite{efgr3}. Moreover, preliminary results in $SU(2)\times SU(2)$
supersymmetric gauge theories, with matter supermultiplets in the $(1,3)$
and $(2,2)^n$ representations, were reported in this symposium by S. Forste,
and will appear in \cite{efgr4}.

\vskip .3in \noindent
{\bf Acknowledgements} \vskip .2in \noindent
The work of SE is supported in part by the BRF - the Basic Research
Foundation.
The work of AG is supported in part by BSF - American-Israel Bi-National
Science Foundation, and by the BRF.
The work of ER is supported in part by BSF and by the BRF.

\vskip .3in \noindent
%\newpage

\end{document}